\documentclass[aip,preprint]{revtex4-1}

\usepackage{graphicx}
\usepackage{float}

\begin{document}

\title{
Effect of ``dipolar-biasing" on the tunability of tunneling magnetoresistance in transition metal oxide systems
}

\author{P. Anil Kumar}
\affiliation{Centre for Advanced Materials, Indian Association for the Cultivation of Science, Kolkata, 700032, India}
\author{D. D. Sarma}
\email[]{Also at Jawaharlal Nehru Centre for Advanced Scientific
Research, Bangalore. Electronic mail: sarma@sscu.iisc.ernet.in}
\affiliation{Solid State and Structural Chemistry Unit, Indian Institute of Science, Bangalore, 560012, India}

\date{\today}

\begin{abstract}
We observe an unusual tunneling magnetoresistance (TMR) phenomenon in a composite of La$_{2/3}$Sr$_{1/3}$MnO$_{3}$ with CoFe$_{2}$O$_{4}$ where the TMR versus applied magnetic field loop suggests a ``negative coercive field". Tracing its origin back to a ``dipolar-biasing" of La$_{2/3}$Sr$_{1/3}$MnO$_{3}$ by CoFe$_{2}$O$_{4}$, we show that the TMR of even a single composite can be tuned continuously so that the resistance peak or the highest sensitivity of the TMR can be positioned anywhere on the magnetic field axis with a suitable magnetic history of the sample. This phenomenon of an unprecedented tunability of the TMR should be present in general in all such composites.
\end{abstract}


\maketitle

Composite structures are known to give rise to distinctly different types of magnetoresistance (MR) like the highly celebrated giant magnetoresistance (GMR)\cite{Baibich19882472,BINASCH19894828} and the presently widely used tunneling magnetoresistance (TMR).\cite{Julliere1975225} Various types of TMR systems have been discussed in the literature, such as the spin-valve type MR where the tunneling is controlled by the magnetic state of the insulating layer.\cite{Sarma2007157205} Also the magnetic tunnel junction type TMR exhibiting MR as large as several hundred percents has been investigated intensely.\cite{Moodera19953273,Parkin2004862,Miao2008246803} TMR is often naturally realized in granular, polycrystalline magnetic metallic materials, where the grain boundaries provide the required insulating barriers. \cite{Coey19983815,Hwang19962041,Li19971124,Coey72734} Deliberately made bulk composites of two polycrystalline entities compacted into a pellet form have also been often used to realize TMR in several cases. \cite{Gerber19976446,Tripathy2007174429} It has been shown that TMR is amenable to tuning, particularly in terms of its magnitude, by different methods, such as grain boundary or interface engineering, suitable composite formation etc. \cite{Moodera19953273,Parkin2004862,Miao2008246803} The present work describes a unique tunability of the functional form of MR(\textit{H}) for the tunneling magnetoresistance of a composite of a soft ferromagnet, La$_{2/3}$Sr$_{1/3}$MnO$_3$ - LSMO with a hard magnetic insulator, CoFe$_2$O$_4$ - CFO. The (LSMO)$_{1-x}$-(CFO)$_x$ (0 $\le$ \textit{x} $\le$ 0.5) powder compacts are prepared by mechanical mixing of LSMO and CFO followed by cold pressing and annealing at 800 $^\circ$C for 1 hour. Annealing at this relatively low temperature gives rise to reasonably hard pellets, while preventing any inter diffusion between LSMO and CFO components across grain boundaries as confirmed by X-ray diffraction and FESEM-EDS analyses. We show that this composite system exhibits an anomalous negative coercive field for the MR(\textit{H}) loop, when compared with the magnetization loop, \textit{M}(\textit{H}). We establish the origin of this important observation as the dipolar coupling of the two components, namely LSMO and CFO. This dipolar coupling leads to unique ways of tuning the TMR behaviour of such a composite material, for example, by allowing one to achieve the highest sensitivity of the tunneling current to small changes in the external magnetic field at any arbitrarily small applied magnetic field.\par The \textit{TMR} response, when viewed over a wide range of the applied magnetic field, for all our samples appears very similar, exhibiting two distinct regimes (i.e. high-field and low-field regimes) depending on the strength of the applied magnetic field.  In this work, we focus on the low-field regime of the \textit{TMR}, shown in Fig. 1 for different samples at 5 K. The black arrows in each figure gives the field sweeping direction. The \textit{TMR} response of pure LSMO in the forward sweep (thick red line in Fig. 1(a)) is zero (highest resistive state) at an applied magnetic field that we denote by \textit{H$^{MR}_C$} throughout this text; this highest resistive state is essentially the reflection of the magnetic coercive field (\textit{H$_C$}, indicated in individual panels of Fig. 1) of the material. Consequently, \textit{H$^{MR}_C$} and \textit{H$_C$} are both positive in the case of pure LSMO. This expected and widely reported equivalence of \textit{H$^{MR}_C$} and \textit{H$_C$}, however, is qualitatively violated for samples with a finite CFO content, as seen in Figs. 1(b)-(f). In each of these cases, the highest resistive state is realised in the wrong direction of the applied magnetic field, as is evident from the figure. For example, \textit{H$_C$} for the sample with 10\% CFO (\textit{x} = 0.1) is at +5 mT, while \textit{H$^{MR}_C$} is at -2.3 mT for the forward sweep of the applied field. This anomalous behaviour of \textit{H$^{MR}_C$}  becomes more prominent with increasing \textit{x}, with \textit{H$^{MR}_C$} moving to more and more negative values, while \textit{H$_C$}, as expected, becoming increasingly positive with increasing CFO content. While exchange coupling between different components of a multi-component system is known to affect the MR of the majority component, an exchange bias coupling between LSMO and CFO is not able to explain the radically anomalous behaviour of MR in these systems, as proven below.
\begin{figure}
\includegraphics{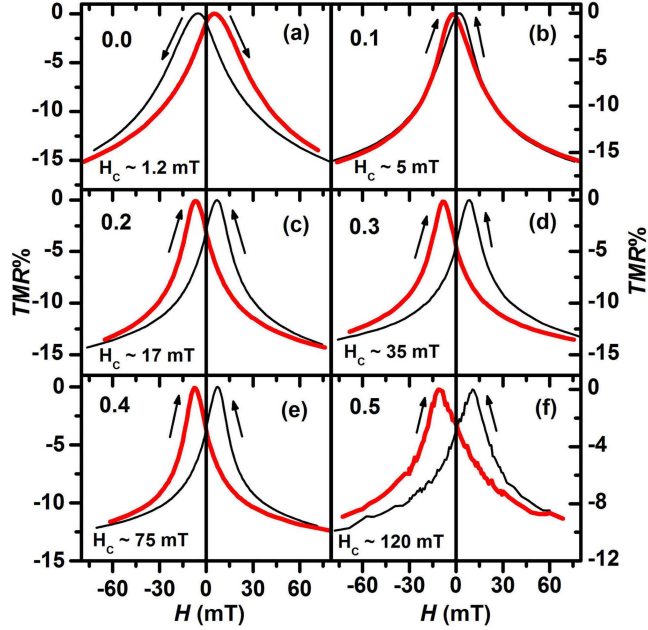}
\caption{\label{Fig.1 } (Color online) The panels present the \textit{TMR} versus field curves for the samples (LSMO)$_{1-x}$-(CFO)$_x$ with (a) \textit{x} = 0.0, (b) \textit{x} = 0.1, (c) \textit{x} = 0.2, (d) \textit{x} = 0.3, (e) \textit{x} = 0.4 and (f)  \textit{x} = 0.5. The arrows in each panel give the direction of the magnetic field sweeping during the measurement.}
\end{figure}
\par In order to explain the observed anomalous TMR behaviour we first look at the hysteresis loops.  The magnetization (\textit{M}) data collected at 5 K over a wide range of \textit{H} are shown for a selected set of samples in Fig. 2(a). \textit{M}(\textit{H}) for each composite, can be grossly described as an average of the significantly different \textit{M}(\textit{H}) of the end-members, LSMO and CFO (Fig. 2(b)). While the agreement between thus calculated average \textit{M}(\textit{H}) and the experimentally measured \textit{M}(\textit{H}) for the composites is very good at high \textit{H}, the agreement is not good at low \textit{H} (Fig. 2(c)).
\begin{figure}
\includegraphics{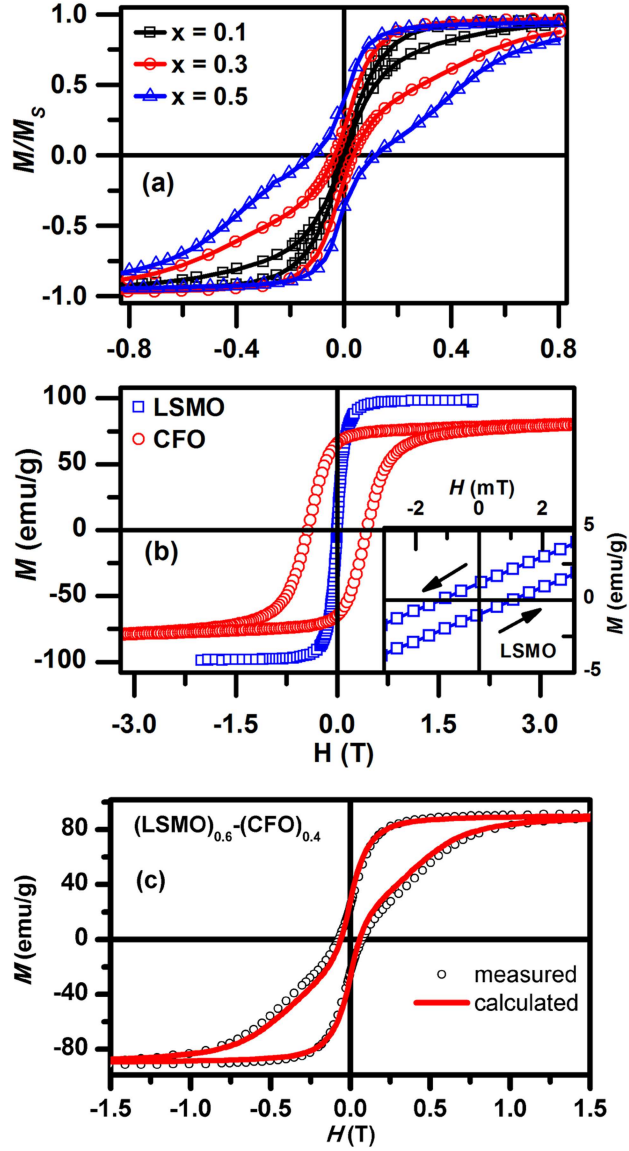}
\caption{\label{Fig. 2 } (Color online) (a) Plots show the magnetization (\textit{M}) versus magnetic field (\textit{H}) data measured at 5 K for the samples (LSMO)$_{1-x}$-(CFO)$_x$  for \textit{x} = 0.1, 0.3 and 0.5. (b) \textit{M}-\textit{H} curves for the samples LSMO and CFO (inset shows the highly magnified \textit{M}-\textit{H} loop for LSMO near its \textit{H$_C$}). (c) The calculated (see text) and experimentally measured \textit{M}-\textit{H} curves for one composition  (LSMO)$_{0.6}$-(CFO)$_{0.4}$.}
\end{figure}
\par We note that any magnetic interaction between LSMO and CFO in these composites, if present, must manifest in deviations from the additive nature of \textit{M}(\textit{H}), described above. In order to probe any subtle signature of such a deviation in \textit{M}(\textit{H}) of these composites, we have subtracted the composition weighted contribution of CFO from the \textit{M}(\textit{H}) of these composites. The remaining part of the \textit{M}(\textit{H}), supposedly representing the extracted \textit{M}(\textit{H}) contribution from the LSMO component of the composite is plotted in Fig. 3(a) for the low field regime (for $x$ = 0.4 composite). This particular representation allows us to focus on subtle influences on the magnetization of LSMO by the presence of CFO for low applied fields. Fig. 3(a) makes it evident that there is a qualitative difference between the thus  obtained \textit{M}(\textit{H}) of the LSMO component in the composite and  that of pure LSMO in its behaviour close to \textit{H} = 0.  Expectedly, the magnetization of pure LSMO becomes zero at a finite positive value of the applied field, defining the coercive field, in the forward sweep of \textit{H} (inset to Fig. 2(b)). In contrast, the \textit{M}(\textit{H}) for the LSMO component of the composite, is zero for a negative value of the applied field in the forward sweep, suggesting a negative coercive field for LSMO in the composite. Similar observations are also found for reverse sweep of \textit{H}. Since the coercive field of a material cannot be negative, it is obvious that LSMO grains in the composites must experience an effective field that is overall positive in spite of a small negative value of the applied field during the forward sweep. This extra field in addition to the external applied field must evidently arise from the magnetic interactions of CFO with LSMO grains.
\begin{figure}
\includegraphics{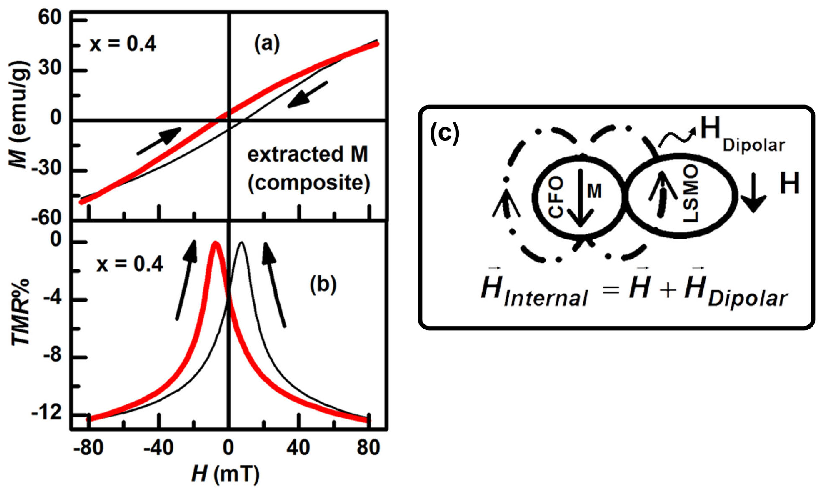}
\caption{\label{Fig. 3} (Color online) (a) Plot shows the resultant \textit{M}-\textit{H} loop, extracted by subtracting the CFO magnetization from that of the composite, for LSMO inside the LSMO-CFO composite with \textit{x} = 0.4. The arrows indicate the field sweeping direction. (b) For comparison the \textit{TMR}-\textit{H} plot for the sample (LSMO)$_{0.6}$-(CFO)$_{0.4}$ is presented. (c) The schematic shows the origin of magnetic dipolar field (\textit{H$_{Dipolar}$}) and its orientation with respect to the external field (\textit{H}).}
\end{figure}
\par It is well known\cite{sattler2010handbook,Gross20035475} that two magnetic grains interact by dipolar fields and this coupling is usually antiferromagnetic, as illustrated in Fig. 3(c). Thus, for all applied magnetic fields lower than the coercive field of CFO during the forward sweep, CFO grains will have a net negative magnetization that will exert an effective positive field on LSMO grains; in particular, for all negative values of the magnetic field in the forward sweep, CFO grains will exert an essentially constant, dipolar field that is positive and therefore, in the opposite direction to the applied external field.  Thus, it becomes evident that the dipolar coupling between CFO and LSMO will help LSMO grains to have a finite positive moment even when the external applied field is still in the opposite direction, as illustrated in Fig. 3(c), for field values smaller than the dipolar field. The effective magnetic field on LSMO, may be represented by $\vec{H}_{Internal} = \vec{H} + \vec{H}_{Dipolar}$ where \textit{H$_{Dipolar}$} is proportional to the magnetization of CFO. Given the large remanence exhibited by the \textit{M}(\textit{H}) loop of CFO, \textit{H$_{Internal}$} is dominated by \textit{H$_{Dipolar}$} for small values of \textit{H}, while at larger values of \textit{H}, \textit{H$_{Internal}$} $\sim$ \textit{H}. Essentially, the magnetization of LSMO is dictated by the field \textit{H$_{Internal}$}, thereby, being controlled by \textit{H$_{Dipolar}$} for \textit{H} $\sim$ 0 and giving rise to the apparently anomalous \textit{H$_C$}, for the LSMO component (Fig. 3(a)). In passing, we note that \textit{H$_{Dipolar}$} for $x$ = 0.4 sample is estimated to be $\sim$ 8.5 mT from the apparent \textit{H$_C$}  of  -7.5 mT (see Fig. 3(a)) and the \textit{H$_C$} ($\sim$1 mT) of pure LSMO (inset to Fig. 2(b)). In Fig. 3(b), we show the \textit{TMR}-\textit{H} loop and compare it to the extracted \textit{M}(\textit{H}) of the LSMO component for the \textit{x} = 0.4 sample. This comparison, showing an excellent agreement between the fields at which \textit{TMR} becomes zero (highest resistive state) and the “coercive field” at which the LSMO magnetization in the composite switches its direction provides a convincing resolution to the apparently anomalous behaviour of \textit{TMR} of all composite samples, shown in Figs. 1(b)-(f). It is appropriate to mention that the dipolar coupling is strongly aniostropic and even its sign might depend on the relative placement of the LSMO grains with respect to the CFO. However, in the present case the LSMO and CFO grains are randomly distributed\cite{APLSI} in the composite pellets and hence the observed dipolar coupling effect is an average effect on the macroscopic behaviour of MR. Nevertheless, we believe that our experimental data cannot be explained by any other mechanism than the dipolar coupling. \par The above discussion, pointing out the importance of the dipolar field in determining the effective field on the LSMO grains, suggests a unique tunability of \textit{TMR} in such composites. As we have already noted, the dipolar field depends on the magnetization of CFO grains which can be easily controlled by the maximum external field the sample is exposed to, or in other words, by the magnetic history of the sample.
\begin{figure}
\includegraphics{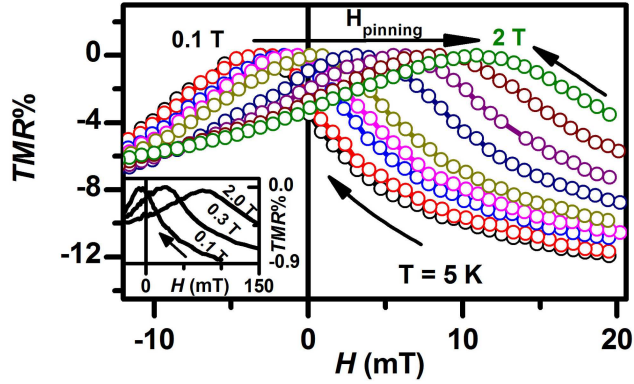}
\caption{\label{Fig. 4 } (Color online) The plot demonstrates the tunability of \textit{TMR} peak position when the \textit{TMR}-\textit{H} loop was measured after exposing the \textit{x} = 0.3 sample to a varying magnetic field \textit{H$_{pinning}$}, from 0.1 T to 2 T (as represented) and limiting the measurement field to the corresponding \textit{H$_{pinning}$}. Inset shows the tunability of \textit{TMR} at 300 K, for a chosen set of \textit{H$_{pinning}$} values as indicated there in.}
\end{figure}
Traversing a minor loop, we can achieve any level of remanent magnetization of CFO grains up to the maximum of saturation remanent magnetization. Having pinned the remanent magnetization of the CFO part, it remains unchanged as long as the sample is not exposed to an external magnetic field comparable to the one used to fix the remanence of CFO grains. Thus, the effective magnetic field on LSMO grains, which is the vector sum of the dipolar field and the external field, is controlled by the original applied field the sample was exposed to, in order to remanently magnetize the CFO grains. This will obviously have two remarkable consequences. First, the low field \textit{TMR} of the composite will be strongly asymmetric with respect to the external magnetic field due to “dipolar biasing” between CFO and LSMO grains and the \textit{TMR} will be strongly dependent on the past magnetic history of the sample. We demonstrate this phenomenon by measuring the low field \textit{TMR} for these samples, after exposing them to various external magnetic fields between 0.1 and 2 T, as illustrated for the \textit{x} = 0.3 sample at 5 K in Fig. 4. Clearly there is a prominent asymmetry and very large tuning of the \textit{TMR} with respect to the applied external magnetic field. 
Therefore, the choice of the original, pinning magnetic field allows us to position the \textit{TMR} in order to achieve the maximum sensitivity at any desired value of the applied field and as close to the \textit{H} = 0 point as may be required for any application. For example, when the external field is changed from \textit{H} = -1 mT to +1 mT, representing a $\Delta$\textit{H} of only 2 mT, our \textit{x} = 0.3 sample, having been exposed to a \textit{H$_{pinning}$} of 0.1 T, shows a 3.8\% change in its resistance, which is well within the detection limit. It should be noted here that the TMR characteristic with zero hysteresis combined with an approximately linear variation of \textit{R} with \textit{H} forms the basis for many low field magnetic sensors.\cite{Baselt1998731} In passing, we note that our preliminary data at \textit{T} = 300 K (inset in Fig. 4) establish the same phenomenon to be operative even up at the room temperature, suggesting possibilities of sensor designs exploring this phenomenon in the future.

Authors thank Department of Science and Technology and Board of Research in Nuclear Sciences, Government of India for supporting this research. Authors thank Proferssors J. M. D. Coey and J. Moodera for useful discussion, Dr R. Rawat for the help with magnetoresistance measurments at UGC-DAE Consortium for Scientific Research, Indore and Mr Koroush Lashgari for his help during SEM analysis. DDS thanks the national J. C. Bose Fellowship.


\begin{thebibliography}{17}%
\makeatletter
\providecommand \@ifxundefined [1]{%
 \@ifx{#1\undefined}
}%
\providecommand \@ifnum [1]{%
 \ifnum #1\expandafter \@firstoftwo
 \else \expandafter \@secondoftwo
 \fi
}%
\providecommand \@ifx [1]{%
 \ifx #1\expandafter \@firstoftwo
 \else \expandafter \@secondoftwo
 \fi
}%
\providecommand \natexlab [1]{#1}%
\providecommand \enquote  [1]{``#1''}%
\providecommand \bibnamefont  [1]{#1}%
\providecommand \bibfnamefont [1]{#1}%
\providecommand \citenamefont [1]{#1}%
\providecommand \href@noop [0]{\@secondoftwo}%
\providecommand \href [0]{\begingroup \@sanitize@url \@href}%
\providecommand \@href[1]{\@@startlink{#1}\@@href}%
\providecommand \@@href[1]{\endgroup#1\@@endlink}%
\providecommand \@sanitize@url [0]{\catcode `\\12\catcode `\$12\catcode
  `\&12\catcode `\#12\catcode `\^12\catcode `\_12\catcode `\%12\relax}%
\providecommand \@@startlink[1]{}%
\providecommand \@@endlink[0]{}%
\providecommand \url  [0]{\begingroup\@sanitize@url \@url }%
\providecommand \@url [1]{\endgroup\@href {#1}{\urlprefix }}%
\providecommand \urlprefix  [0]{URL }%
\providecommand \Eprint [0]{\href }%
\providecommand \doibase [0]{http://dx.doi.org/}%
\providecommand \selectlanguage [0]{\@gobble}%
\providecommand \bibinfo  [0]{\@secondoftwo}%
\providecommand \bibfield  [0]{\@secondoftwo}%
\providecommand \translation [1]{[#1]}%
\providecommand \BibitemOpen [0]{}%
\providecommand \bibitemStop [0]{}%
\providecommand \bibitemNoStop [0]{.\EOS\space}%
\providecommand \EOS [0]{\spacefactor3000\relax}%
\providecommand \BibitemShut  [1]{\csname bibitem#1\endcsname}%
\let\auto@bib@innerbib\@empty
\bibitem [{\citenamefont {Baibich}\ \emph {et~al.}(1988)\citenamefont
  {Baibich}, \citenamefont {Broto}, \citenamefont {Fert}, \citenamefont
  {Van~Dau}, \citenamefont {Petroff}, \citenamefont {Eitenne}, \citenamefont
  {Creuzet}, \citenamefont {Friedrich},\ and\ \citenamefont
  {Chazelas}}]{Baibich19882472}%
  \BibitemOpen
  \bibfield  {author} {\bibinfo {author} {\bibfnamefont {M.~N.}\ \bibnamefont
  {Baibich}}, \bibinfo {author} {\bibfnamefont {J.~M.}\ \bibnamefont {Broto}},
  \bibinfo {author} {\bibfnamefont {A.}~\bibnamefont {Fert}}, \bibinfo {author}
  {\bibfnamefont {F.~N.}\ \bibnamefont {Van~Dau}}, \bibinfo {author}
  {\bibfnamefont {F.}~\bibnamefont {Petroff}}, \bibinfo {author} {\bibfnamefont
  {P.}~\bibnamefont {Eitenne}}, \bibinfo {author} {\bibfnamefont
  {G.}~\bibnamefont {Creuzet}}, \bibinfo {author} {\bibfnamefont
  {A.}~\bibnamefont {Friedrich}}, \ and\ \bibinfo {author} {\bibfnamefont
  {J.}~\bibnamefont {Chazelas}},\ }\href@noop {} {\bibfield  {journal}
  {\bibinfo  {journal} {Phys. Rev. Lett.}\ }\textbf {\bibinfo {volume} {61}},\
  \bibinfo {pages} {2472--2475} (\bibinfo {year} {1988})}\BibitemShut {NoStop}%
\bibitem [{\citenamefont {Binasch}\ \emph {et~al.}(1989)\citenamefont
  {Binasch}, \citenamefont {Grunberg}, \citenamefont {Saurenbach},\ and\
  \citenamefont {Zinn}}]{BINASCH19894828}%
  \BibitemOpen
  \bibfield  {author} {\bibinfo {author} {\bibfnamefont {G.}~\bibnamefont
  {Binasch}}, \bibinfo {author} {\bibfnamefont {P.}~\bibnamefont {Grunberg}},
  \bibinfo {author} {\bibfnamefont {F.}~\bibnamefont {Saurenbach}}, \ and\
  \bibinfo {author} {\bibfnamefont {W.}~\bibnamefont {Zinn}},\ }\href@noop {}
  {\bibfield  {journal} {\bibinfo  {journal} {Phys. Rev. B}\ }\textbf {\bibinfo
  {volume} {39}},\ \bibinfo {pages} {4828--4830} (\bibinfo {year}
  {1989})}\BibitemShut {NoStop}%
\bibitem [{\citenamefont {Julliere}(1975)}]{Julliere1975225}%
  \BibitemOpen
  \bibfield  {author} {\bibinfo {author} {\bibfnamefont {M.}~\bibnamefont
  {Julliere}},\ }\href@noop {} {\bibfield  {journal} {\bibinfo  {journal}
  {Phys. Lett. A}\ }\textbf {\bibinfo {volume} {54}},\ \bibinfo {pages} {225 --
  226} (\bibinfo {year} {1975})}\BibitemShut {NoStop}%
\bibitem [{\citenamefont {Sarma}\ \emph {et~al.}(2007)\citenamefont {Sarma},
  \citenamefont {Ray}, \citenamefont {Tanaka}, \citenamefont {Kobayashi},
  \citenamefont {Fujimori}, \citenamefont {Sanyal}, \citenamefont
  {Krishnamurthy},\ and\ \citenamefont {Dasgupta}}]{Sarma2007157205}%
  \BibitemOpen
  \bibfield  {author} {\bibinfo {author} {\bibfnamefont {D.~D.}\ \bibnamefont
  {Sarma}}, \bibinfo {author} {\bibfnamefont {S.}~\bibnamefont {Ray}}, \bibinfo
  {author} {\bibfnamefont {K.}~\bibnamefont {Tanaka}}, \bibinfo {author}
  {\bibfnamefont {M.}~\bibnamefont {Kobayashi}}, \bibinfo {author}
  {\bibfnamefont {A.}~\bibnamefont {Fujimori}}, \bibinfo {author}
  {\bibfnamefont {P.}~\bibnamefont {Sanyal}}, \bibinfo {author} {\bibfnamefont
  {H.~R.}\ \bibnamefont {Krishnamurthy}}, \ and\ \bibinfo {author}
  {\bibfnamefont {C.}~\bibnamefont {Dasgupta}},\ }\href@noop {} {\bibfield
  {journal} {\bibinfo  {journal} {Phys. Rev. Lett.}\ }\textbf {\bibinfo
  {volume} {98}},\ \bibinfo {pages} {157205} (\bibinfo {year}
  {2007})}\BibitemShut {NoStop}%
\bibitem [{\citenamefont {Moodera}\ \emph {et~al.}(1995)\citenamefont
  {Moodera}, \citenamefont {Kinder}, \citenamefont {Wong},\ and\ \citenamefont
  {Meservey}}]{Moodera19953273}%
  \BibitemOpen
  \bibfield  {author} {\bibinfo {author} {\bibfnamefont {J.~S.}\ \bibnamefont
  {Moodera}}, \bibinfo {author} {\bibfnamefont {L.~R.}\ \bibnamefont {Kinder}},
  \bibinfo {author} {\bibfnamefont {T.~M.}\ \bibnamefont {Wong}}, \ and\
  \bibinfo {author} {\bibfnamefont {R.}~\bibnamefont {Meservey}},\ }\href@noop
  {} {\bibfield  {journal} {\bibinfo  {journal} {Phys. Rev. Lett.}\ }\textbf
  {\bibinfo {volume} {74}},\ \bibinfo {pages} {3273 -- 3276} (\bibinfo {year}
  {1995})}\BibitemShut {NoStop}%
\bibitem [{\citenamefont {Parkin}\ \emph {et~al.}(2004)\citenamefont {Parkin},
  \citenamefont {Kaiser}, \citenamefont {Panchula}, \citenamefont {Rice},
  \citenamefont {Hughes}, \citenamefont {Samant},\ and\ \citenamefont
  {Yang}}]{Parkin2004862}%
  \BibitemOpen
  \bibfield  {author} {\bibinfo {author} {\bibfnamefont {S.~S.~P.}\
  \bibnamefont {Parkin}}, \bibinfo {author} {\bibfnamefont {C.}~\bibnamefont
  {Kaiser}}, \bibinfo {author} {\bibfnamefont {A.}~\bibnamefont {Panchula}},
  \bibinfo {author} {\bibfnamefont {P.~M.}\ \bibnamefont {Rice}}, \bibinfo
  {author} {\bibfnamefont {B.}~\bibnamefont {Hughes}}, \bibinfo {author}
  {\bibfnamefont {M.}~\bibnamefont {Samant}}, \ and\ \bibinfo {author}
  {\bibfnamefont {S.-H.}\ \bibnamefont {Yang}},\ }\href@noop {} {\bibfield
  {journal} {\bibinfo  {journal} {Nat. Mater.}\ }\textbf {\bibinfo {volume}
  {3}},\ \bibinfo {pages} {862--867} (\bibinfo {year} {2004})}\BibitemShut
  {NoStop}%
\bibitem [{\citenamefont {Miao}\ \emph {et~al.}(2008)\citenamefont {Miao},
  \citenamefont {Park}, \citenamefont {Moodera}, \citenamefont {Seibt},
  \citenamefont {Eilers},\ and\ \citenamefont {Munzenberg}}]{Miao2008246803}%
  \BibitemOpen
  \bibfield  {author} {\bibinfo {author} {\bibfnamefont {G.~X.}\ \bibnamefont
  {Miao}}, \bibinfo {author} {\bibfnamefont {Y.~J.}\ \bibnamefont {Park}},
  \bibinfo {author} {\bibfnamefont {J.~S.}\ \bibnamefont {Moodera}}, \bibinfo
  {author} {\bibfnamefont {M.}~\bibnamefont {Seibt}}, \bibinfo {author}
  {\bibfnamefont {G.}~\bibnamefont {Eilers}}, \ and\ \bibinfo {author}
  {\bibfnamefont {M.}~\bibnamefont {Munzenberg}},\ }\href@noop {} {\bibfield
  {journal} {\bibinfo  {journal} {Phys. Rev. Lett.}\ }\textbf {\bibinfo
  {volume} {100}},\ \bibinfo {pages} {246803} (\bibinfo {year}
  {2008})}\BibitemShut {NoStop}%
\bibitem [{\citenamefont {Coey}\ \emph
  {et~al.}(1998{\natexlab{a}})\citenamefont {Coey}, \citenamefont {Berkowitz},
  \citenamefont {Balcells}, \citenamefont {Putris},\ and\ \citenamefont
  {Barry}}]{Coey19983815}%
  \BibitemOpen
  \bibfield  {author} {\bibinfo {author} {\bibfnamefont {J.~M.~D.}\
  \bibnamefont {Coey}}, \bibinfo {author} {\bibfnamefont {A.~E.}\ \bibnamefont
  {Berkowitz}}, \bibinfo {author} {\bibfnamefont {L.~l.}\ \bibnamefont
  {Balcells}}, \bibinfo {author} {\bibfnamefont {F.~F.}\ \bibnamefont
  {Putris}}, \ and\ \bibinfo {author} {\bibfnamefont {A.}~\bibnamefont
  {Barry}},\ }\href@noop {} {\bibfield  {journal} {\bibinfo  {journal} {Phys.
  Rev. Lett.}\ }\textbf {\bibinfo {volume} {80}},\ \bibinfo {pages} {3815 --
  3818} (\bibinfo {year} {1998}{\natexlab{a}})}\BibitemShut {NoStop}%
\bibitem [{\citenamefont {Hwang}\ \emph {et~al.}(1996)\citenamefont {Hwang},
  \citenamefont {Cheong}, \citenamefont {Ong},\ and\ \citenamefont
  {Batlogg}}]{Hwang19962041}%
  \BibitemOpen
  \bibfield  {author} {\bibinfo {author} {\bibfnamefont {H.~Y.}\ \bibnamefont
  {Hwang}}, \bibinfo {author} {\bibfnamefont {S.-W.}\ \bibnamefont {Cheong}},
  \bibinfo {author} {\bibfnamefont {N.~P.}\ \bibnamefont {Ong}}, \ and\
  \bibinfo {author} {\bibfnamefont {B.}~\bibnamefont {Batlogg}},\ }\href@noop
  {} {\bibfield  {journal} {\bibinfo  {journal} {Phys. Rev. Lett.}\ }\textbf
  {\bibinfo {volume} {77}},\ \bibinfo {pages} {2041 -- 2044} (\bibinfo {year}
  {1996})}\BibitemShut {NoStop}%
\bibitem [{\citenamefont {Li}\ \emph {et~al.}(1997)\citenamefont {Li},
  \citenamefont {Gupta}, \citenamefont {Xiao},\ and\ \citenamefont
  {Gong}}]{Li19971124}%
  \BibitemOpen
  \bibfield  {author} {\bibinfo {author} {\bibfnamefont {X.~W.}\ \bibnamefont
  {Li}}, \bibinfo {author} {\bibfnamefont {A.}~\bibnamefont {Gupta}}, \bibinfo
  {author} {\bibfnamefont {G.}~\bibnamefont {Xiao}}, \ and\ \bibinfo {author}
  {\bibfnamefont {G.~Q.}\ \bibnamefont {Gong}},\ }\href@noop {} {\bibfield
  {journal} {\bibinfo  {journal} {Appl. Phys. Lett.}\ }\textbf {\bibinfo
  {volume} {71}},\ \bibinfo {pages} {1124 -- 1126} (\bibinfo {year}
  {1997})}\BibitemShut {NoStop}%
\bibitem [{\citenamefont {Coey}\ \emph
  {et~al.}(1998{\natexlab{b}})\citenamefont {Coey}, \citenamefont {Berkowitz},
  \citenamefont {l.~Balcells}, \citenamefont {Putris},\ and\ \citenamefont
  {Parker}}]{Coey72734}%
  \BibitemOpen
  \bibfield  {author} {\bibinfo {author} {\bibfnamefont {J.~M.~D.}\
  \bibnamefont {Coey}}, \bibinfo {author} {\bibfnamefont {A.~E.}\ \bibnamefont
  {Berkowitz}}, \bibinfo {author} {\bibfnamefont {L.}~\bibnamefont
  {l.~Balcells}}, \bibinfo {author} {\bibfnamefont {F.~F.}\ \bibnamefont
  {Putris}}, \ and\ \bibinfo {author} {\bibfnamefont {F.~T.}\ \bibnamefont
  {Parker}},\ }\href@noop {} {\bibfield  {journal} {\bibinfo  {journal} {Appl.
  Phys. Lett.}\ }\textbf {\bibinfo {volume} {72}},\ \bibinfo {pages} {734 --
  736} (\bibinfo {year} {1998}{\natexlab{b}})}\BibitemShut {NoStop}%
\bibitem [{\citenamefont {Gerber}\ \emph {et~al.}(1997)\citenamefont {Gerber},
  \citenamefont {Milner}, \citenamefont {Groisman}, \citenamefont {Karpovsky},
  \citenamefont {Gladkikh},\ and\ \citenamefont {Sulpice}}]{Gerber19976446}%
  \BibitemOpen
  \bibfield  {author} {\bibinfo {author} {\bibfnamefont {A.}~\bibnamefont
  {Gerber}}, \bibinfo {author} {\bibfnamefont {A.}~\bibnamefont {Milner}},
  \bibinfo {author} {\bibfnamefont {B.}~\bibnamefont {Groisman}}, \bibinfo
  {author} {\bibfnamefont {M.}~\bibnamefont {Karpovsky}}, \bibinfo {author}
  {\bibfnamefont {A.}~\bibnamefont {Gladkikh}}, \ and\ \bibinfo {author}
  {\bibfnamefont {A.}~\bibnamefont {Sulpice}},\ }\href@noop {} {\bibfield
  {journal} {\bibinfo  {journal} {Phys. Rev. B}\ }\textbf {\bibinfo {volume}
  {55}},\ \bibinfo {pages} {6446 -- 6452} (\bibinfo {year} {1997})}\BibitemShut
  {NoStop}%
\bibitem [{\citenamefont {Tripathy}, \citenamefont {Adeyeye},\ and\
  \citenamefont {Shannigrahi}(2007)}]{Tripathy2007174429}%
  \BibitemOpen
  \bibfield  {author} {\bibinfo {author} {\bibfnamefont {D.}~\bibnamefont
  {Tripathy}}, \bibinfo {author} {\bibfnamefont {A.~O.}\ \bibnamefont
  {Adeyeye}}, \ and\ \bibinfo {author} {\bibfnamefont {S.}~\bibnamefont
  {Shannigrahi}},\ }\href@noop {} {\bibfield  {journal} {\bibinfo  {journal}
  {Phys. Rev. B}\ }\textbf {\bibinfo {volume} {76}},\ \bibinfo {pages} {174429}
  (\bibinfo {year} {2007})}\BibitemShut {NoStop}%
\bibitem [{\citenamefont {Sattler}(2010)}]{sattler2010handbook}%
  \BibitemOpen
  \bibfield  {author} {\bibinfo {author} {\bibfnamefont {K.}~\bibnamefont
  {Sattler}},\ }\href@noop {} {\emph {\bibinfo {title} {Handbook of
  Nanophysics: Nanoparticles and Quantum Dots}}},\ Handbook of Nanophysics\
  (\bibinfo  {publisher} {Taylor \& Francis},\ \bibinfo {year}
  {2010})\BibitemShut {NoStop}%
\bibitem [{\citenamefont {Gross}\ \emph {et~al.}(2003)\citenamefont {Gross},
  \citenamefont {Diehl}, \citenamefont {Beverly}, \citenamefont {Richman},\
  and\ \citenamefont {Tolbert}}]{Gross20035475}%
  \BibitemOpen
  \bibfield  {author} {\bibinfo {author} {\bibfnamefont {A.}~\bibnamefont
  {Gross}}, \bibinfo {author} {\bibfnamefont {M.}~\bibnamefont {Diehl}},
  \bibinfo {author} {\bibfnamefont {K.}~\bibnamefont {Beverly}}, \bibinfo
  {author} {\bibfnamefont {E.}~\bibnamefont {Richman}}, \ and\ \bibinfo
  {author} {\bibfnamefont {S.}~\bibnamefont {Tolbert}},\ }\href {\doibase
  {10.1021/jp034240n}} {\bibfield  {journal} {\bibinfo  {journal} {{J. Phys.
  Chem. B}}\ }\textbf {\bibinfo {volume} {{107}}},\ \bibinfo {pages}
  {{5475--5482}} (\bibinfo {year} {{2003}})}\BibitemShut {NoStop}%
\bibitem [{APL()}]{APLSI}%
  \BibitemOpen
  \href@noop {} {}\bibinfo {note} {See supplementary material at [URL will be
  inserted by AIP] for SEM image and EDS elemental mapping of the
  composite.}\BibitemShut {Stop}%
\bibitem [{\citenamefont {Baselt}\ \emph {et~al.}(1998)\citenamefont {Baselt},
  \citenamefont {Lee}, \citenamefont {Natesan}, \citenamefont {Metzger},
  \citenamefont {Sheehan},\ and\ \citenamefont {Colton}}]{Baselt1998731}%
  \BibitemOpen
  \bibfield  {author} {\bibinfo {author} {\bibfnamefont {D.~R.}\ \bibnamefont
  {Baselt}}, \bibinfo {author} {\bibfnamefont {G.~U.}\ \bibnamefont {Lee}},
  \bibinfo {author} {\bibfnamefont {M.}~\bibnamefont {Natesan}}, \bibinfo
  {author} {\bibfnamefont {S.~W.}\ \bibnamefont {Metzger}}, \bibinfo {author}
  {\bibfnamefont {P.~E.}\ \bibnamefont {Sheehan}}, \ and\ \bibinfo {author}
  {\bibfnamefont {R.~J.}\ \bibnamefont {Colton}},\ }\href@noop {} {\bibfield
  {journal} {\bibinfo  {journal} {Biosensors \& Bioelectronics}\ }\textbf
  {\bibinfo {volume} {13}},\ \bibinfo {pages} {731--739} (\bibinfo {year}
  {1998})}\BibitemShut {NoStop}%
\end{thebibliography}
%

\end{document}